\documentstyle[aps,pre,epsf,twocolumn,amssymb]{revtex}

\begin{document}

\twocolumn[
\hsize\textwidth\columnwidth\hsize\csname @twocolumnfalse\endcsname

\title{L\'{e}vy-flight Spreading of Epidemic Processes leading to Percolating
Clusters}
\author{H.K.\ Janssen$^1$, K.\ Oerding$^1$, F.\ van Wijland$^2$, H.J.\ 
Hilhorst$^2$}
\address{$^1$Institut f\"{u}r Theoretische Physik III, 
Heinrich-Heine-Universit\"{a}t,
\\
40225 D\"{u}sseldorf, Germany}
\address{$^{2}$Laboratoire de Physique Th\'{e}orique et Hautes Energies \thanks{
{\small
Laboratoire~associ\'{e}~au~Centre~National~de~la~Recherche~Scientifique~-URA~D00
63}},\\
Universit\'{e} de Paris-Sud,\\ 91405 Orsay cedex, France}
\date{\today}
\maketitle

\widetext
\begin{abstract}
We consider two stochastic processes, the Gribov process and the
general epidemic process, that describe the spreading of an
infectious disease. In contrast to the usually assumed case of
short-range infections that lead, at the critical point, to
directed and isotropic percolation respectively, we consider
long-range infections with a probability distribution decaying in $
d $ dimensions with the distance as $1/R^{d+\sigma}$. By means of
Wilson's momentum shell renormalization-group recursion relations,
the critical exponents characterizing the growing fractal clusters
are calculated to first order in an $\varepsilon$-expansion. It is
shown that the long-range critical behavior changes continuously to
its short-range counterpart for a decay exponent of the infection
$\sigma =\sigma _{c}>2$.
\end{abstract}

\narrowtext
]


\section{Introduction}

\label{Intro}

\subsection{Epidemic Processes}

\label{EpPro}

The investigation of the formation and of the properties of random
structures has been an exciting topic in statistical physics for many years.
In the case that the formation of such structures obeys local rules, these
processes can often be expressed in the language of population growth. It is
well known that two special growth processes called (in the language of a
disease) simple epidemic with recovery (Gribov process \cite
{GrassSu78,GrassTo79}, also known in elementary particle physics as Reggeon
field theory \cite{Gri67,GM68,Mo78}, the stochastic version of Schl\"{o}gls
first reaction \cite{Janssen18,Schl72}) and epidemic with removal (general
epidemic process (GEP) \cite{Mol77,Bai75,Murray8}) lead to random structures
with the properties of percolation clusters: directed percolation \cite
{BrHa57,CardySugar8,Obukhov} in the first case and isotropic percolation
(for a recent overview see \cite{BuHa96}) in the last one \cite
{Grass83,Janssen28,CardyGrassberger8}. These stochastic processes describe
the essential features of a vast number of growth phenomena of populations
of infected individuals near their extinction threshold and are relevant to a
wide range of models in physics, chemistry, biology, and sociology. The
transition between survival and extinction of such a growing population is a
nonequilibrium continuous phase transition phenomenon and is characterized
by universal scaling laws.

The Gribov process with short-range infection belongs to the universality
class of local growth processes with absorbing states \cite
{Janssen18,Grass82} such as the contact process \cite{Ha74,Li85,JeDi94} and
certain cellular automata \cite{Kinzel83,Kinzel85}. This universality class
is characterized by the following four principles:

\begin{enumerate}
\item  Infection of susceptible (``birth'') and spontaneous annihilation
(``recovery'') of infected individuals.

\item  Interaction (``saturation'') between the infected individuals.

\item  Diffusion (``spreading'') of the disease in a $d$-dimensional space.

\item  The state without infected individuals is absorbing.
\end{enumerate}

To model these principles in a universal form, we use from the
beginning a mesoscopic picture in which all microscopic length- and
time-scales are considered as very short. Thus we take a continuum
approach with the density $n\left( {\bf x},t\right) $ of the
infected individuals (the ills) at time $ t $ as an order
parameter. Note that the spontaneous annihilation of the ills makes
it possible to avoid complications arising in the case of only
diffusion controlled reactions which need creation and destruction
operators as order parameters for a correct description.

The Langevin equation is constructed in accordance with the four principles
as
\begin{equation}
\partial _{t}n=\lambda \nabla ^{2}n+R\left[ n\right] n+\zeta ,  \label{Lang}
\end{equation}
where $\zeta \left( {\bf x},t\right) $ denotes a Gaussian Markovian
noise with short range correlations which has to vanish if $n\left(
{\bf x} ,t\right) =0$ to model the absorbing state and the reaction
rate $R\left[ n
\right] $ models birth, recovery and saturation. In a low density expansion
we may set \cite{Janssen18}
\begin{eqnarray}
R\left[ n\right] &=&-\lambda \left( \tau +\frac{g}{2}n\right) ,
\label{RateG} \\
\langle \zeta \left( {\bf x},t\right) \zeta \left( {\bf x}^{\prime
},t^{\prime }\right) \rangle &=&\lambda g^{\prime }n\left( {\bf x},t\right)
\delta \left( {\bf x-x}^{\prime }\right) \delta \left( t-t^{\prime }\right) .
\label{LForce}
\end{eqnarray}

In contrast to the Gribov process (GP), the general epidemic process (GEP)
introduces besides the susceptibles, $S$, who can catch the disease, and the
infectives or ills, $I$, who have the disease and can transmit it, as a
third class the removed, $R$, namely those who have had the disease and are
now immune or death. Thus the first principle above is to be modified to

\begin{enumerate}
\item[1.']  Infection of susceptible and spontaneous annihilation but
without recovery of susceptible individuals.
\end{enumerate}

Therefore the reaction rate now also depends on the number of the removed
individuals, introducing a memory term into the process. Because this term
is the leading one in the long-time and large-distance limit we now have 
\cite{Janssen28}
\begin{eqnarray}
R\left[ n\right] &=&-\lambda \left( \tau +gm\right) ,  \label{RateE} \\
m\left( {\bf x},t\right) &=&\lambda \int_{-\infty }^{t}dt^{\prime }\,n\left(
{\bf x},t^{\prime }\right) .  \label{Remove}
\end{eqnarray}

In a microscopic realization of the Gribov process a single species of
(quasi)particles, the $I$'s, is introduced. The $I$'s represent the infected
individuals (sites of a lattice). They perform simple random walks and
undergo the following ``chemical'' reaction scheme built from reversible
branching and irreversible spontaneous annihilation:
\[
\text{Gribov Process}\left\{
\begin{array}{l}
I\leftrightarrow 2I \\
I\rightarrow \emptyset
\end{array}
\right. .
\]
Above some value of the branching rate of the reaction
$I\rightarrow 2I$, the stationary state has a finite density of
$I$'s. As the branching rate goes down to a threshold value, the
stationary state density of $I$'s goes continuously to zero, which
is an absorbing state below the threshold. This threshold value
corresponds to the critical point $\tau =\tau _{c}$ in Eq.~(
\ref{RateG}). $\tau _{c}=0$ as long as one neglects fluctuation
contributions.

A microscopic model which belongs to the universality class of the general
epidemic process involves the three species $S$, $I$, and $R$ specified
above. Only the infected individuals $I$ are mobile. A susceptible $S$ may
be contaminated, but the infected $I$'s may become spontaneously immune:
\[
\text{General Epidemic Process}\left\{
\begin{array}{l}
S+I\rightarrow 2I \\
I\rightarrow R
\end{array}
\right.
\]
The history of this model goes back to 1927 when it was first
introduced in the mathematical biology literature \cite{Murray8}.
Here of course the stationary state is $I$-free. In this model the
key parameter is the initial density of $S$'s, denoted by $\rho $:
depending on the value $\rho $ with respect to a threshold value
$\rho _{c}$, the infected individuals may either start to
proliferate as a solitary wave before dying out in a finite system,
which occurs for $\rho >\rho _{c}$ ($\tau <\tau _{c}$ in Eq.~(\ref
{RateE})), or their number decreases from the outset, which takes
place for $
\rho <\rho _{c}$ ($\tau >\tau _{c}$).

\subsection{L\'{e}vy-flight Infections}

\label{LevFl}

In the standard version of the epidemic models the susceptible individuals
can become contaminated by already infected {\em neighboring} individuals.
At the same time infected individuals are subject to spontaneous healing or
immunization processes.

In realistic situations the infection can be also long-ranged. This may be
e.g.\ by a disease in an orchard where flying parasites contaminate the
trees practically instantaneous in a widespread manner if the timescale of
the flights of the parasites is much shorter as the mesoscopic timescale of
the epidemic process itself. Thus following a suggestion of Mollison \cite
{Mol77}, Grassberger \cite{Grassberger} introduced a variation of the
epidemic processes with infection probability distributions $P\left(
R\right) $ which decays with the distance $R$ as a power-law like
\begin{equation}
P(R)\propto \frac{1}{R^{d+\sigma }},\qquad \text{for\quad }R\rightarrow
\infty .  \label{LongRange}
\end{equation}
We will in general denote such long-range distributions as
L\'{e}vy-flights although a true L\'{e}vy-flight is defined by its
Fourier transform as $
\tilde{P}\left( {\bf q}\right) \propto \exp \left( -b\left| {\bf q}\right|
^{\sigma }\right) $ \cite{MoWe79}, and then only L\'{e}vy-exponents with $
0<\sigma \leq 2$ give rise to positive distributions \cite{Bo59}. The
infection rate in the Langevin equation (\ref{Lang}) is now given by
\begin{equation}
\left. \frac{\partial n\left( {\bf x},t\right) }{\partial t}\right| _{\text{
inf}}=\int d^{d}x^{\prime }\,P\left( \left| {\bf x-x}^{\prime }\right|
\right) n\left( {\bf x}^{\prime },t\right)  \label{InfRate}
\end{equation}
After Fourier transformation of this equation and after a small momentum
expansion that is relevant in our mesoscopic consideration we get
\begin{equation}
\left. \frac{\partial \tilde{n}\left( {\bf q},t\right) }{\partial t}\right|
_{\text{inf}}=\left( p_{0}-p_{2}q^{2}+p_{\sigma }q^{\sigma }+o\left(
q^{2},q^{\sigma }\right) \right) \tilde{n}\left( {\bf q},t\right)
\label{FouInfR}
\end{equation}
where the analytical terms stem from the short-range part of
$P\left( R\right) $ and the nonanalytical ones arise from the
power-law decay (\ref {LongRange}). The constant $p_{0}$ is
included in the reaction rate as a negative (``birth'')
contribution to $\tau $ whereas $p_{2}$ yields a diffusional term.
Naively the parameter $p_{\sigma }$ is relevant or irrelevant in
the long wave-length limit if $\sigma $ is smaller or bigger than
$2$ respectively, and this fact has mislead some authors to neglect
this term from the outset if $\sigma >2$. But this naive
(``Gaussian'') argumentation may be wrong in an interacting theory
because the critical behavior is in general determined by an
nontrivial fixed point of a renormalization group transformation.
To decide which one of the terms in Eq.~(\ref{FouInfR}) are
relevant, one has to compare with the scaling behavior of the
Fourier transformed susceptibility $\chi \left( {\bf q} ,\omega
\right) \propto q^{2-\bar{\eta}}$. If $\sigma <2-\bar{\eta}$, the
parameter $p_{\sigma }$ is a relevant perturbation and must be
included in a renormalization group procedure. Prominent examples
of systems with $\bar{
\eta}<0$ are $\phi ^{3}$-models as e.g.\ the Yang-Lee-singularity model \cite
{Janssen98}. In all these cases $p_{\sigma }$ is relevant also for $\sigma
>2 $ and cannot be neglected.

In the following we define $\sigma =2\left( 1-\alpha \right) $ and the
diffusion term in Eq.~(\ref{Lang}) is now completed by a term proportional
to $\propto q^{2(1-\alpha )}n\left( q,t\right) $. In real space we write the
completed Langevin equation as
\begin{equation}
\partial _{t}n=\lambda \left[ 1-\frac{v}{2\alpha }\left( \left( -\nabla
^{2}\right) ^{-\alpha }+1\right) \right] \nabla ^{2}n+R\left[ n\right]
n+\zeta ,  \label{LRLang}
\end{equation}
and the gradient-terms should be only considered (in Fourier space) up to a
cutoff $\Lambda $ that we have set to $1$. Then stability of this terms
against inhomogeneous perturbation is guaranteed if $v\geq 0$.

Grassberger \cite{Grassberger} reported new critical exponents for $\alpha
>0 $ from a 1-loop calculation that contain some numerical errors. These
exponents are discontinuous in the limit $\alpha \rightarrow +0$ if
one assumes irrelevance of the new terms for $\alpha <0$. In this
paper we will reconsider the problem and show that the full range
of values $\alpha >\bar{
\eta}_{SR}/2$ lead to new critical behavior. Here $\bar{\eta}_{SR}<0$ is the
anomalous susceptibility exponent of the epidemic models with
short-range infection. We will show that the critical exponents
change {\em continuously }at the boundary $2\alpha
=\bar{\eta}_{SR}$ from long-range to short-range behavior.

We remark that the interest in reaction-diffusion problems
involving particles that perform L\'{e}vy flights is not new. In
the physics literature, they have most recently arisen as follows.
Particles performing simple random walks are subject to the
reactions $A+B\rightarrow \emptyset $ and $A+A\rightarrow \emptyset
$ in the presence of a quenched velocity field
\cite{ZumofenKlafter8}. The effect of the quenched velocity field
is then to enhance diffusion in such a way that the effective
action of the velocity field is reproduced if L\'{e}vy flights are
substituted for the simple random walk motion. In the above
mentioned reactions the time decay of the particle density is
algebraic with an exponent related to that of the step length
distribution of the L\'{e}vy flights defined in
Eq.~(\ref{LongRange}). These results have been confirmed by
several renormalization group calculations
\cite{Oerding8,DeemPark8}.

\section{The Gribov process with L\'{e}vy-Flights}

\label{GribPro}

\subsection{Renormalization Group Analysis}

\label{GrRen}

In order to develop the renormalization group analysis we recast the
Langevin equation (\ref{LRLang}) as a dynamic functional \cite
{Janssen76,DeDominicis,Ja92,Janssen18}
\begin{eqnarray}
{\cal J}\left[ \tilde{s},s\right] &=&\int d^{d}x\,dt\,\tilde{s}\left\{
\partial _{t}+\lambda \frac{g}{2}\left( s-\tilde{s}\right) \right.  \nonumber
\\
&&\left. +\lambda \left[ \tau -\nabla ^{2}+\frac{v}{2\alpha }\left( \left(
-\nabla ^{2}\right) ^{1-\alpha }+\nabla ^{2}\right) \right] \right\} s
\label{GDynFu}
\end{eqnarray}
where $\tilde{s}$ is Martin-Siggia-Rose response field~\cite{MSR}.
We note that he dynamic functional can also be derived
using the methods developed in \cite{Doi,GrassScheun,Peliti8} from
a microscopic master equation. By a suitable rescaling of the
density $n\varpropto s$, the constant $g^{\prime }$ in
Eq.~(\ref{LForce}) is made equal to $g.$ The dynamic functional
(\ref{GDynFu}) is then symmetric in the absorbing phase under the
exchange $s\left( {\bf x},t\right) \leftrightarrow -\tilde{s}
\left( {\bf x},-t\right) $. All correlation and response functions can be
calculated as functional integrals with weight $\exp \left( -{\cal J}\right)
$ in a perturbation expansion involving the propagator (the unperturbed
response function)
\begin{equation}
G_{0}\left( {\bf q},t\right) =\Theta \left( t\right) \exp \left\{
-\lambda
\left[ \tau +q^{2}+\frac{v}{2\alpha }\left( q^{-2\alpha }-1\right) q^{2}
\right] t\right\}  \label{Prop}
\end{equation}
as a function of momentum ${\bf q}$ and time $t$. This propagator
guarantees stability for all $\alpha $ as long as $\tau \geq 0$,
$v\geq 0$, and $ q=\left| {\bf q}\right| \leq 1$. For simplicity we
have set the momentum cut-off $\Lambda =1$.

To study the critical behavior of this system near the critical
point we use Wilson's renormalization procedure. We introduce the
usual coarse graining parameter $b>1$ and split the fields $s$ and
$\tilde{s}$ into components which are non zero on the momentum
shell $\Omega _{b}=\left\{ {\bf q} |1/b\leq |{\bf q}|\leq 1\right\}
$ and components defined on the complement of $\Omega _{b}$, the
latter being denoted by $s^{<}$ and ${\tilde{s}}^{<}$. We integrate
out the short scale degrees of freedom in the weight $\exp
\left( -{\cal J}\right) $, that is, those defined on $\Omega _{b}$, and
rescale the fields according to
\begin{eqnarray}
s\left( {\bf x},t\right)  &\rightarrow &s^{\prime }\left(
b^{-1}{\bf x} ,b^{-2-\zeta }t\right) =b^{\left( d+\gamma \right)
/2}s^{<}\left( {\bf x} ,t\right) ,  \nonumber \\
\tilde{s}\left( {\bf x},t\right)  &\rightarrow &\tilde{s}^{\prime }\left(
b^{-1}{\bf x},b^{-2-\zeta }t\right) =b^{\left( d+\gamma \right)
/2}\tilde{s}
^{<}\left( {\bf x},t\right) .  \label{RescaleG}
\end{eqnarray}
Renormalized parameters $\tau ^{\prime }$, $v^{\prime }$, and $g^{\prime }$
are defined in such a way that the coarse grained functional looks like the
old one. The one-loop calculation is standard and does not present any
technical difficulties. For infinitesimal renormalization transformation
with $b-1\ll 1$ we obtain
\begin{eqnarray}
&&i\omega +\lambda \left[ \tau ^{\prime }+q^{2}+\frac{v^{\prime
}}{2\alpha }
\left( q^{-2\alpha }-1\right) q^{2}\right]   \nonumber \\
&=&i\omega b^{-\gamma }\left[ 1-\frac{u}{4\left( 1+\tau \right)
^{2}}\ln b
\right]   \nonumber \\
&&+\lambda b^{2+\zeta -\gamma }\left[ \tau +\frac{u}{2\left( 1+\tau
\right) }
\ln b\right]   \nonumber \\
&&+\lambda q^{2}b^{\zeta -\gamma }\left[ 1+\frac{v}{2\alpha }\left(
b^{2\alpha }q^{-2\alpha }-1\right) \vspace{0.15cm}\right.   \nonumber \\
&&\qquad \qquad \qquad \left. -\frac{uK\left( v\right) }{8\left( 1+\tau
\right) ^{2}}\ln b\right]   \label{ElimG1}
\end{eqnarray}
and
\begin{equation}
u^{\prime }=ub^{4-d+2\zeta -3\gamma }\left[ 1-\frac{2u}{\left( 1+\tau
\right) ^{2}}\ln b\right] .  \label{ElimG2}
\end{equation}
Here $\omega $ is the frequency, $u=S_{d}g^{2}/2$ with $S_{d}$ the
surface of the unit sphere in $d$ dimensions divided by $\left(
2\pi \right) ^{d}$ , and $K\left( v\right) =1-cv$, where $c$ is
an uninteresting positive constant. Note that the calculation
neglects terms of order $O\left( u^{2}\right) $ but is exact (for the
coarse graining method we have used) with respect to the parameter $v$.
By comparison of the terms $\propto
\omega $ and $q^{2}$ in Eq.~(\ref{ElimG1}) we get the
Wilson-functions
\begin{eqnarray}
\gamma  &=&-\frac{u}{4\left( 1+\tau \right) ^{2}}+O\left( u^{2}\right) ,
\nonumber \\
\bar{\gamma} &=&\gamma -\zeta =v-\frac{uK\left( v\right) }{8\left( 1+\tau
\right) ^{2}}+O\left( u^{2}\right) .  \label{GParFu}
\end{eqnarray}

We use $l=\ln b$ as the flow parameter of the renormalization transformation
that yields then from the other terms in Eqs.~(\ref{ElimG1},\ref{ElimG2})
the flow equations
\begin{eqnarray}
\frac{d\tau }{dl} &=&\left[ \left( 2-\bar{\gamma}\right) \tau +\frac{u}{
2\left( 1+\tau \right) }+O\left( u^{2}\right) \right] ,  \label{GtaFlow} \\
\frac{du}{dl} &=&\left[ 4-d-\gamma -2\bar{\gamma}-\frac{2u}{\left( 1+\tau
\right) ^{2}}+O\left( u^{2}\right) \right] ,  \label{GuFlow} \\
\frac{dv}{dl} &=&\left( 2\alpha -\bar{\gamma}\right) v.  \label{GvFlow}
\end{eqnarray}
The last equation is exact since the operator $\tilde{s}\left( -\nabla
^{2}\right) ^{1-\alpha }s$ in the dynamic functional (\ref{GDynFu}) is not
renormalized. The reason is that the renormalization procedure can only
generate contributions that are analytic in the momenta.

In order to study the fixed point structure we find it useful to
introduce $\bar{\varepsilon}=4\left( 1-\alpha \right) -d=\varepsilon
-4\alpha $. Near the
fixed points we linearize the flow equation (\ref{GtaFlow}) for the relevant
variable $\tau $ about $\tau _{c}=\tau _{\ast }$ as $d\tau /dl\thickapprox
\nu ^{-1}\left( \tau -\tau ^{\ast }\right) $. The flow equations (\ref
{GtaFlow}-\ref{GvFlow}) have, besides the trivial short-range
Gaussian fixed point $\left( \tau ^{\ast },u^{\ast },v^{\ast
}\right) =\left( 0,0,0\right)$, stable for $\alpha <0$ and $d>4$,
and the trivial L\'{e}vy-Gaussian fixed point $\left( \tau ^{\ast
},u^{\ast },v^{\ast }\right) =\left( 0,0,2\alpha
\right) $ with $\bar{\eta}_{LG}=\bar{\gamma}^{\ast }=2\alpha ,$ $z=2+\zeta
^{\ast }=2\left( 1-\alpha \right) $, stable for $d>4(1-\alpha )$ and $\alpha
>0$, two non trivial fixed points. The first one is the already known
short-range directed percolation fixed point $\left( \tau ^{\ast },u^{\ast
},v^{\ast }\right) =\left( -\varepsilon /3,2\varepsilon /3,0\right) +O\left(
\varepsilon ^{2}\right) $, with $\bar{\eta}_{DP}=\bar{\gamma}^{\ast
}=-\varepsilon /12+O\left( \varepsilon ^{2}\right) ,$ $z_{DP}=2+\zeta ^{\ast
}=2-\varepsilon /12+O\left( \varepsilon ^{2}\right) ,$ $\nu
_{DP}^{-1}=2-\varepsilon /4+O\left( \varepsilon ^{2}\right) $, stable for $
\varepsilon >0,$ $\alpha <-\varepsilon /24+O\left( \varepsilon ^{2}\right) $
. The second one is the new L\'{e}vy-directed-percolation fixed point
\begin{eqnarray}
u^{\ast } &=&\frac{4}{7}\bar{\varepsilon}+O\left( \bar{\varepsilon}
^{2}\right) ,\quad   \nonumber \\
v^{\ast } &=&\frac{28\alpha
+\bar{\varepsilon}}{14+c\bar{\varepsilon}} +O\left(
\bar{\varepsilon}^{2}\right) ,  \nonumber \\
\tau ^{\ast } &=&\tau _{c}=-\frac{\bar{\varepsilon}}{7\left( 1-\alpha
\right) }+O\left( \bar{\varepsilon}^{2}\right)   \label{LGfix}
\end{eqnarray}
which is stable for $-\varepsilon /24+O\left( \varepsilon ^{2}\right)
<\alpha <\varepsilon /4$. We obtain for this fixed point the critical
exponents
\begin{eqnarray}
\eta _{LDP} &=&\gamma ^{\ast }=-\frac{\bar{\varepsilon}}{7}+O\left( \bar{
\varepsilon}^{2}\right) ,  \label{LGet} \\
\bar{\eta}_{LDP} &=&\bar{\gamma}^{\ast }=2\alpha ,  \label{LGetq} \\
z_{LDP} &=&2+\eta _{LDP}-\bar{\eta}_{LDP}  \nonumber \\ &=&2\left(
1-\alpha \right) -\frac{\bar{\varepsilon}}{7}+O\left( \bar{
\varepsilon}^{2}\right) ,  \label{LGz} \\
\nu _{LDP}^{-1} &=&2\left( 1-\alpha \right) -\frac{2\bar{\varepsilon}}{7}
+O\left( \bar{\varepsilon}^{2}\right) .  \label{LGnu}
\end{eqnarray}
Note that at all fixed points $u^{\ast }$ and $v^{\ast }$ are non-negative
as they should for stability of the theory.

\begin{figure}[hp]
\vspace*{-53mm}
\epsfxsize=400pt
\hspace*{-25mm}\epsfbox{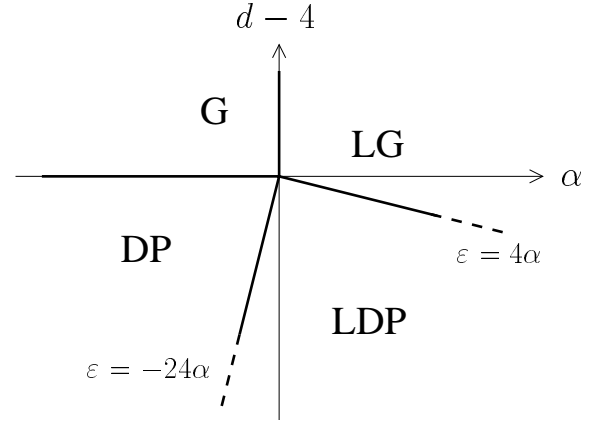}
\vspace*{-85mm}
\caption{Stability regions of the Gribov process with long range
spreading in the $(\alpha,d=4-\varepsilon)$ plane. G, LG, DP and LDP
indicate the respective stability regions of the short
range Gaussian, L\'evy Gaussian, short range directed percolation and 
L\'evy directed percolation fixed point}
\label{abb1}
\end{figure}

We have depicted in Fig.\thinspace \ref{abb1} the stability regions for
each of the above fixed points in the $(\alpha , d)$ plane. A glance on
the exact flow equation (\ref{GvFlow}) of the parameter $v$ shows
that the boundary between the domains of attraction of the directed
percolation fixed point and the L\'{e}vy-directed-percolation fixed
point is given exactly by $\bar{\eta}
_{DP}=\bar{\eta}_{LDP}=2\alpha $. At this boundary all exponents change {\em
continuously} upon varying the parameter $\alpha $ from their values at the
directed percolation fixed point to those of the
L\'{e}vy-directed-percolation fixed point and vice versa.

\subsection{Scaling analysis}

\label{GrScal}

In the following we consider the scaling behavior of two key quantities: the
time dependent order parameter (the density of infected individuals) $\rho
\left( t\right) =\langle s\left( {\bf x},t\right) \rangle _{\rho _{0}}$ for $
t>0$ if the initial state at time $t=0$ is prepared with a
homogeneous initial density $\rho _{0}$ , and the response function
$\chi ({\bf x} ,t)=\langle s\left( {\bf x},t\right) \tilde{s}\left(
{\bf 0},0\right)
\rangle $ that yields the density of infected individuals after the epidemic
is initialized by a pointlike source at $t=0$ and ${\bf x}=0$. Here we are
interested in the L\'{e}vy-flight case only, thus we will disregard the
subscripts at all critical exponents. We set $\tau -\tau _{c}\rightarrow
\tau $ in the following.

>From the rescaling Eqs.~(\ref{RescaleG}) we get at the L\'{e}vy fixed point
(which is approached for $b\gg 1$) the relationship
\begin{equation}
\rho \left( t,\tau ,\rho _{0}\right) =b^{-\left( d+\eta \right) /2}\rho
\left( b^{-z}t,b^{1/\nu }\tau ,b^{\left( d-\eta \right) /2}\rho _{0}\right) ,
\label{GrOrd}
\end{equation}
where the critical exponents are displayed by Eqs.\ (\ref{LGet}-\ref{LGnu}).
The scaling of the initial density is easily found by noting that $\rho
_{0}\ $arises in the Langevin equation (\ref{LRLang}) as an additive source
term $q({\bf x},t)=\rho _{0}\delta \left( t\right) $ that
translates in the dynamic functional ${\cal J}$,
Eq.~(\ref{GDynFu}), to a further additive contribution $\int
d^{d}x\,\rho _{0}\tilde{s}({\bf x},0)$ from which one directly
reads off the scaling behavior of $\rho _{0}$ if one knows that $
\tilde{s}({\bf x},0)$ scales as $\tilde{s}({\bf x},t)$ \cite{WiOeHi98}. At
criticality, for $\tau =0$, one obtains from Eq.~(\ref{GrOrd}) that the
order parameter first increases in a universal initial time regime \cite
{JaSS89,Ja92} as
\begin{equation}
\rho \left( t,\rho _{0}\right) \propto \rho _{0}t^{\theta }  \label{GInit}
\end{equation}
where the universal initial time exponent $\theta $ is given by
\begin{equation}
\theta =-\frac{\eta }{z}=\frac{{\bar{\varepsilon}}}{7\sigma }{+O}\left( \bar{
\varepsilon}^{2}\right) ,  \label{Gtheta}
\end{equation}
and we have set $\sigma =2\left( 1-\alpha \right) $. Then, after some
crossover time, the order parameter decreases as
\begin{eqnarray}
\rho \left( t\right)  &\propto &t^{-\left( d+\eta \right) /2z},\qquad
\nonumber \\
\text{with}\qquad \frac{d+\eta }{2z} &=&1-\frac{3{\bar{\varepsilon}}}{
7\sigma }{+O}\left( \bar{\varepsilon}^{2}\right) .  \label{Gdecay}
\end{eqnarray}
In the non absorbing stationary state, i.e.$\ $for $\tau <0$ and $
t\rightarrow \infty $, the order parameter behaves as
\begin{equation}
\rho _{stat}\left( \tau \right) \propto \left| \tau \right| ^{\beta },
\label{GEqSt}
\end{equation}
where the order parameter exponent $\beta $ is found as
\begin{equation}
\beta =\frac{\nu \left( d+\eta \right) }{2}=1-\frac{2{\bar{\varepsilon}}}{
7\sigma }{+O}\left( \bar{\varepsilon}^{2}\right) .  \label{Gbeta}
\end{equation}
Finally, at the critical dimension $d=2\sigma =4\left( 1-\alpha \right) $,
the scaling behavior is mean-field like with logarithmic corrections,
\begin{eqnarray}
\rho \left( t\right)  &\propto &\rho _{0}\ln ^{1/7}t\qquad   \nonumber \\
&&\text{in the initial time regime at criticality,}  \nonumber \\
\rho \left( t\right)  &\propto &\frac{\ln ^{3/7}t}{t}\qquad \quad   \nonumber
\\
&&\text{in the long time regime at criticality,}  \nonumber \\
\rho _{\text{stat}}\left( \tau \right)  &\propto &\left| \tau \right| \ln
^{2/7}\left( \frac{1}{|\tau|} \right) \qquad   \nonumber \\
&&\text{in the stationary state,}  \label{GlogKorr}
\end{eqnarray}
which we mention for completeness.

The scaling behavior of the response function is given by
\begin{equation}
\chi \left( {\bf x},t\right) =b^{-(d+\eta )}\chi \left( b^{-1}{\bf x}
,b^{-z}t,b^{1/\nu }\tau \right) .  \label{Gchi}
\end{equation}
First we read off the correlation lengths for space and time and the
corresponding exponents as
\begin{eqnarray}
\xi _{\perp } &\propto &\left| \tau \right| ^{-\nu _{\perp }}\qquad
\nonumber \\
\text{with}\qquad \nu _{\perp } &=&\nu =\frac{1}{\sigma }+\frac{2{\bar{
\varepsilon}}}{7\sigma ^{2}}{+O}\left( \bar{\varepsilon}^{2}\right) ,
\label{Gorth} \\
\xi _{\parallel } &\propto &\left| \tau \right| ^{-\nu _{\parallel }}\qquad
\nonumber \\
\text{with}\qquad \nu _{\parallel } &=&z\nu =1+\frac{{\bar{\varepsilon}}}{
7\sigma }{+O}\left( \bar{\varepsilon}^{2}\right) .  \label{Glong}
\end{eqnarray}
At criticality, $\tau =0$, we have
\begin{equation}
\chi \left( {\bf x},t\right) =t^{-\left( d+\eta \right) /z}{\cal F}\left(
{\bf x/}t^{1/z}\right)   \label{Gcritchi}
\end{equation}
with a universal scaling function ${\cal F}\left( x\right) .$ This relation
shows that the density of infected individuals as the result from a
pointlike seed dies out with an exponent $(d+\eta )/z=2\beta /\nu z$.
Comparing with the general decay law (\ref{Gdecay}), we find that the
probability to find after a time $t$ an infected individual if there was a
pointlike seed at time $t=0$ decays with an exponent $\beta /\nu z$. The
Fourier transformed susceptibility at the critical point scales as
\begin{equation}
\chi \left( {\bf q},\omega \right) =q^{-2+\bar{\eta}}\widetilde{{\cal F}}
\left( \omega {\bf /}q^{z}\right) \propto q^{-\sigma }.  \label{GchiFou}
\end{equation}
As the last scaling exponent that can be deduced in the usual way from the
given critical exponents we present the fractal dimension of the clusters of
the infected individuals:
\begin{eqnarray}
d_{f} &=&d-\frac{\beta }{\nu }=\frac{d-\eta }{2}  \nonumber \\
&=&\sigma -\frac{3\bar{\varepsilon}}{7}{+O}\left( \bar{\varepsilon}
^{2}\right) .  \label{Gfracd}
\end{eqnarray}
We note that the values of the exponents $\nu _{\parallel }$
(\ref{Gorth}), $
\nu _{\perp }$ (\ref{Glong}), $\beta $ (\ref{Gbeta}), and $d_{f}$ (\ref
{Gfracd}) that we have found are different from those given by
Grassberger \cite{Grassberger}. The values of all the exponents
changes {\em continuously }at the stability boundary $\sigma
=2\left( 1-\alpha \right) =2-
\bar{\eta}_{DP}=2+\varepsilon /12+O\left( \varepsilon ^{2}\right) $ to their
short-range directed percolation values.

\subsection{Comparison with existing simulations}

\label{GrSim}

In a recent letter Albano~\cite{Albano} has presented a numerical
study of one-dimensional branching and annihilating random walks
(BARW) in which the individuals perform L\'{e}vy flights. For
Brownian particles in $d<2$ dimensions the BARW (which is defined
by the equations $A\rightarrow (m+1)A$ and $A+A\rightarrow
\emptyset $) is known to belong for $m$ odd to the universality
class of directed percolation~\cite{CardyTaeuber1,CardyTaeuber2}
. For $d>2$ the systems shows a phase transition at zero branching rate
which can be described by mean field exponents. If the random walk is
replaced by L\'{e}vy flights, noise becomes irrelevant above $d_{\sigma
}=\sigma $.

Albano has investigated the behavior of the critical exponents for $m=1$ as
a function of $\sigma $, for $0.25\leq \sigma \leq 11$. His results are
summarized in the following table:
\begin{equation}
\begin{array}{||c|c|c|c||}
\hline\hline
\sigma & z & \eta & z-\sigma -\eta \\ \hline\hline
2 & 1.590 & -0.482 & 0.072 \\ \hline
1.5 & 1.585 & -0.483 & 0.568 \\ \hline
1 & 1.583 & -0.489 & 1.073 \\ \hline
0.75 & 1.581 & -0.512 & 1.343 \\ \hline
0.5 & 1.575 & -0.553 & 1.628 \\ \hline
0.25 & 1.569 & -0.574 & 1.893 \\ \hline\hline
\end{array}
\label{table}
\end{equation}
The critical dimension $d_{\sigma }$ is lower than $1$ for $\sigma \in
\left\{ 0.25,0.5,0.75\right\} $. (For $\sigma =0.25$ even the critical
dimension of the L\'{e}vy-flight directed percolation $d_{c}=4(1-\alpha
)=2\sigma $ is lower than $1$). Therefore the phase transition should occur
at zero branching rate, and critical exponents should be the mean field ones
($\beta _{\text{mf}}=1$, $z_{\text{mf}}=\sigma $ and $\eta _{\text{mf}}=0$);
this is clearly not the case for the exponents given in table~(\ref{table}).
The hyperscaling relation $z=\sigma +\eta $ Eqs.~(\ref{LGetq},\ref{LGz}),
which we have shown to hold in any dimension, is violated as $\sigma $ is
decreased. These facts cast doubt on the reliability of the simulations
performed in \cite{Albano}. A possible explanation is to be found in the
L\'{e}vy flight generation procedure. Indeed the author uses a distribution
Eq.~(\ref{LongRange}) truncated at some distance cut-off, the effect of
which is to produce an effectively {\em short range} motion. This
interpretation is confirmed by the slow variation of the exponents as a
function of $\sigma $, their values remaining close to that of directed
percolation with simple random walk displacement.

\section{The General Epidemic Process with L\'{e}vy-Flights}

\label{GEP}

\subsection{Renormalization Group Analysis}

\label{GEPRen}

The renormalization group analysis of the GEP is performed analogously to
the corresponding analysis of the Gribov process presented in section 2. The
Langevin equation for the GEP (\ref{LRLang}), where now the reaction rate is
given by Eq.~(\ref{RateE}), is recast in the dynamic functional \cite
{Janssen28}:
\begin{eqnarray}
{\cal J}\left[ \tilde{s},s\right]  &=&\int d^{d}x\,dt\,\tilde{s}\left\{
\partial _{t}+\lambda gS-\lambda \frac{g}{2}\tilde{s}\right.   \nonumber \\
&&\left. +\lambda \left[ \tau -\nabla ^{2}+\frac{v}{2\alpha }\left( \left(
-\nabla ^{2}\right) ^{1-\alpha }+\nabla ^{2}\right) \right] \right\} s.
\label{EDynFu}
\end{eqnarray}
The field $S\left( {\bf x},t\right) =\lambda \int_{-\infty
}^{t}dt^{\prime }\,s\left( {\bf x},t^{\prime }\right) $, a rescaled
form of Eq.~(\ref{Remove} ), introduces a memory term in the
dynamics. In analogy to the Gribov process we have rescaled the
fields so that $g^{\prime }=g$. The dynamic functional
(\ref{EDynFu}) is then symmetric under the exchange $S\left( {\bf
x},-t\right) \leftrightarrow -\tilde{s}\left( {\bf x},t\right) $,
or $ s\left( {\bf x},-t\right) \leftrightarrow \partial
_{t}\tilde{s}\left( {\bf x },t\right) $ \cite{Janssen28}. From
this symmetry follows that we only need to consider one coupling
coefficient $g$ for the two interaction terms in $ {\cal J}$. The
perturbation expansion involves also the propagator displayed in
Eq.~(\ref{Prop}), and $v\geq 0$ is needed for stability. We
integrate out the short scale degrees of freedom in the weight
$\exp \left( -{\cal J}
\right) $, and rescale now the fields according to
\begin{eqnarray}
s\left( {\bf x},t\right)  &\rightarrow &s^{\prime }\left(
b^{-1}{\bf x} ,b^{-2-\zeta }t\right) =b^{\left( d+2+\gamma \right)
/2}s^{<}\left( {\bf x} ,t\right) ,  \nonumber \\
\tilde{s}\left( {\bf x},t\right)  &\rightarrow &\tilde{s}^{\prime }\left(
b^{-1}{\bf x},b^{-2-\zeta }t\right) =b^{\left(
d-2+\tilde{\gamma}\right) /2}
\tilde{s}^{<}\left( {\bf x},t\right) .  \label{RescaleGEP}
\end{eqnarray}
Note that from the exchange symmetry follows exactly
\begin{equation}
\zeta =\frac{\gamma -\tilde{\gamma}}{2}  \label{Zeta}
\end{equation}
The renormalized parameters $\tau ^{\prime }$, $v^{\prime }$, and $g^{\prime
}$ are now defined by the coarse graining equations calculated to one-loop
order with $b-1\ll 1$
\begin{eqnarray}
&&i\omega +\lambda \left[ \tau ^{\prime }+q^{2}+\frac{v^{\prime
}}{2\alpha }
\left( q^{-2\alpha }-1\right) q^{2}\right]   \nonumber \\
&=&i\omega b^{-\left( \gamma +\tilde{\gamma}\right) /2}\left[
1-\frac{3u}{ 4\left( 1+\tau \right) ^{3}}\ln b\right]   \nonumber
\\ &&+\lambda b^{2-\tilde{\gamma}}\left[ \tau +\frac{u}{2\left(
1+\tau \right)
^{2}}\ln b\right]   \nonumber \\
&&+\lambda q^{2}b^{-\tilde{\gamma}}\left[ 1+\frac{v}{2\alpha
}\left( b^{2\alpha }q^{-2\alpha }-1\right) \vspace{0.15cm}\right.
\nonumber \\ &&\qquad \qquad \qquad \left. -\frac{\left( d-2\right)
uK\left( v\right) }{ 4d\left( 1+\tau \right) ^{2}}\ln b\right]
\label{ElimGEP1}
\end{eqnarray}
and
\begin{equation}
u^{\prime }=ub^{6-d-3\tilde{\gamma}}\left[ 1-\frac{4u}{\left( 1+\tau \right)
^{3}}\ln b\right] ,  \label{ElimGEP2}
\end{equation}
where $u=S_{d}g^{2}$, and $K\left( v,\tau =0\right) =1-cv$, and $c$ is an
uninteresting positive constant. The Wilson functions and the
renormalization group equations follows now as
\begin{eqnarray}
\gamma +\tilde{\gamma} &=&-\frac{3u}{2\left( 1+\tau \right) ^{3}}+O\left(
u^{2}\right) ,  \label{GEPParFu1} \\
\tilde{\gamma} &=&v-\frac{\left( d-2\right) uK\left( v\right) }{4d\left(
1+\tau \right) ^{2}}+O\left( u^{2}\right) ,  \label{GEPParFu2} \\
\frac{d\tau }{dl} &=&\left[ \left( 2-\tilde{\gamma}\right) \tau +\frac{u}{
2\left( 1+\tau \right) ^{2}}+O\left( u^{2}\right) \right] ,
\label{GEPtaFlow} \\
\frac{du}{dl} &=&\left[ 6-d-3\tilde{\gamma}-\frac{4u}{\left( 1+\tau \right)
^{3}}+O\left( u^{2}\right) \right] u,  \label{GEPuFlow} \\
\frac{dv}{dl} &=&\left( 2\alpha -\tilde{\gamma}\right) v,  \label{GEPvFlow}
\end{eqnarray}

The last equation is exact.

In order to study the fixed point structure we find it useful here to
introduce $\bar{\varepsilon}=6\left( 1-\alpha \right) -d=\varepsilon
-6\alpha $. The flow equations (\ref{GEPtaFlow}-\ref{GEPvFlow}) have,
besides the trivial short-range Gaussian fixed point $\left( \tau
^{\ast },u^{\ast },v^{\ast }\right) =\left( 0,0,0\right) $, stable
for $\alpha <0$ and $d>6$, and the trivial L\'{e}vy-Gaussian fixed
point $\left( \tau ^{\ast },u^{\ast },v^{\ast }\right) =\left(
0,0,2\alpha \right) $ with $\bar{\eta}
_{LG}=\tilde{\gamma}^{\ast }=2\alpha ,$ $\eta _{LG}=\gamma _{\ast }=-2\alpha
$, $z=2+\zeta ^{\ast }=2\left( 1-\alpha \right) $, stable for $d>6\left(
1-\alpha \right) $ and $\alpha >0$, two non trivial fixed points. The first
one is the already known short-range-GEP fixed point (where the static
exponents are known from undirected percolation) $\left( \tau ^{\ast
},u^{\ast },v^{\ast }\right) =\left( -\varepsilon /7,2\varepsilon
/7,0\right) +O\left( \varepsilon ^{2}\right) $, with $\bar{\eta}_{GEP}=\bar{
\gamma}^{\ast }=-\varepsilon /21+O\left( \varepsilon ^{2}\right) ,$ $\eta
_{GEP}=\gamma ^{\ast }=-8\varepsilon /21+O\left( \varepsilon ^{2}\right) ,$ $
z_{GEP}=2+\zeta ^{\ast }=2-\varepsilon /6+O\left( \varepsilon ^{2}\right) ,$
$\nu _{GEP}^{-1}=2-5\varepsilon /21+O\left( \varepsilon ^{2}\right) $,
stable for $\varepsilon >0,$ $\alpha <-\varepsilon /42+O\left( \varepsilon
^{2}\right) $. The second one is the new L\'{e}vy-GEP fixed point
\begin{eqnarray}
u^{\ast } &=&\frac{1}{4}\bar{\varepsilon}+O\left( \bar{\varepsilon}
^{2}\right) ,\quad   \nonumber \\
v^{\ast } &=&\frac{96\alpha \left( 1-\alpha \right) +\left( 2-3\alpha
\right) \bar{\varepsilon}}{48\left( 1-\alpha \right) +c\left( 2-3\alpha
\right) \bar{\varepsilon}}+O\left( \bar{\varepsilon}^{2}\right) ,  \nonumber
\\
\tau ^{\ast } &=&\tau _{c}=-\frac{\bar{\varepsilon}}{16\left( 1-\alpha
\right) }+O\left( \bar{\varepsilon}^{2}\right)   \label{LGEPfix}
\end{eqnarray}
which is stable for $-\varepsilon /42+O\left( \varepsilon ^{2}\right)
<\alpha <\varepsilon /6$. We obtain for this fixed point the critical
exponents
\begin{eqnarray}
\eta _{LGEP} &=&\gamma ^{\ast }=-2\alpha -\frac{3\bar{\varepsilon}}{8}
+O\left( \bar{\varepsilon}^{2}\right) ,  \label{LGEPet} \\
\bar{\eta}_{LGEP} &=&\bar{\gamma}^{\ast }=2\alpha ,  \label{LGEPetq} \\
z_{LGEP} &=&2+\zeta ^{\ast }=2\left( 1-\alpha \right) -\frac{3\bar{
\varepsilon}}{16}+O\left( \bar{\varepsilon}^{2}\right) ,  \label{LGEPz} \\
\nu _{LGEP}^{-1} &=&2\left( 1-\alpha \right) -\frac{\bar{\varepsilon}}{4}
+O\left( \bar{\varepsilon}^{2}\right) .  \label{LGEPnu}
\end{eqnarray}

\begin{figure}
\vspace*{-53mm}
\epsfxsize=400pt
\hspace*{-25mm}\epsfbox{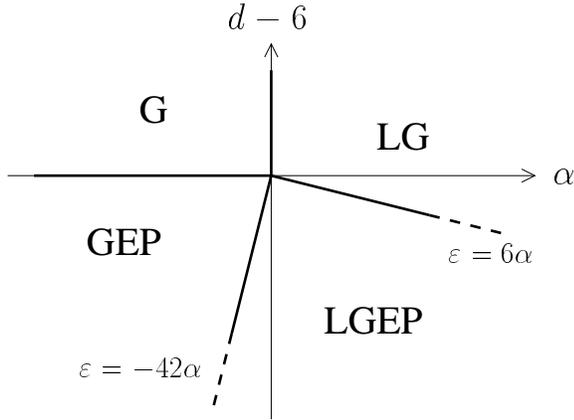}
\vspace*{-85mm}
\caption{Stability regions of the general epidemic process with long
range spreading in the $(\alpha, d=6-\varepsilon)$ plane. G, LG, GEP
and LGEP represent the stability regions of the short range Gaussian,
L\'evy Gaussian, short range percolation and L\'evy percolation
fixed point, respectively.}
\label{abb2}
\end{figure}

Note again that at all fixed points $u^{\ast }$ and $v^{\ast }$ are
non-negative as they should for stability of the theory. We have
depicted in Fig.\thinspace \ref{abb2} the stability regions for each of the
above fixed points in the $\left( \alpha ,d\right) $ plane. Now a
glance at the exact flow equation (\ref{GEPvFlow}) of the parameter
$v$ shows that the boundary between the domains of attraction of
the short-range-GEP fixed point and the L\'{e}vy-GEP
fixed point is
given exactly by $\bar{\eta}_{GEP}=\bar{\eta}
_{LGEP}=2\alpha $. At this boundary all exponents
are again {\em continuous} functions of the parameter $\alpha$.

\subsection{Scaling analysis}

\label{GEPScal}

At first we consider the scaling behavior of the time dependent density of
infected individuals $\rho \left( t\right) =\langle s\left( {\bf x},t\right)
\rangle _{\rho _{0}}$ for $t>0$ if the initial state at time $t=0$ is
prepared with a homogeneous initial density $\rho _{0}$. We are interested
in the L\'{e}vy-flight case only, thus we will disregard the subscripts at
all critical exponents. Again we set $\tau -\tau _{c}\rightarrow \tau $.

Here we find from the rescaling Eqs.~(\ref{RescaleGEP}) at the L\'{e}vy
fixed point (which is approached for $b\gg 1$) the relationship
\begin{equation}
\rho \left( t,\tau ,\rho _{0}\right) =b^{-\left( d+2+\eta \right) /2}\rho
\left( b^{-z}t,b^{1/\nu }\tau ,b^{\left( d+2-\bar{\eta}\right) /2}\rho
_{0}\right) ,  \label{GEPdens}
\end{equation}
where the critical exponents are displayed by Eqs.\
(\ref{LGEPet}-\ref {LGEPnu}). The scaling of the initial density is
also found by adding a source term $q({\bf x},t)=\rho _{0}\delta
\left( t\right) $ to the Langevin equation (\ref{LRLang}) that
translates in the dynamic functional ${\cal J}$ ,
Eq.~(\ref{EDynFu}), to the additive contribution $\int d^{d}x\,\rho
_{0}
\tilde{s}({\bf x},0)$ from which one reads off the scaling behavior of $\rho
_{0}$ since $\tilde{s}({\bf x},0)$ scales as $\tilde{s}({\bf x}
,t)$. At criticality, for $\tau =0$, it follows from Eq.~(\ref{GEPdens})
that the infection density first increases in the universal initial time
regime as
\begin{equation}
\rho \left( t,\rho _{0}\right) \propto \rho _{0}t^{\theta }  \label{GEPInit}
\end{equation}
where the universal initial time exponent $\theta $ is given by
\begin{eqnarray}
\theta  &=&-\frac{\eta +\bar{\eta}}{2z}=\frac{\sigma }{z}-1  \nonumber \\
&=&\frac{3{\bar{\varepsilon}}}{16\sigma }{+O}\left(
\bar{\varepsilon}
^{2}\right) .  \label{GEPtheta}
\end{eqnarray}
We have set $\sigma =2\left( 1-\alpha \right) $.

As an order parameter we consider the density of the removed (immune)
individuals (\ref{Remove}) namely
\begin{eqnarray}
\bar{\rho}\left( t,\tau ,\rho _{0}\right)  &\propto &\langle S\left( {\bf x}
,t\right) \rangle _{\rho _{0}}  \nonumber \\
&=&\lambda \int_{-\infty }^{t}dt^{\prime }\,\langle s\left( {\bf x},t\right)
\rangle _{\rho _{0}}.  \label{GEPOrd}
\end{eqnarray}
The scaling properties of this order parameter are determined from Eq.~(\ref
{GEPdens}) by
\begin{equation}
\bar{\rho}\left( t,\tau ,\rho _{0}\right) =b^{-\left( d-2+\bar{\eta}\right)
/2}\bar{\rho}\left( b^{-z}t,b^{1/\nu }\tau ,b^{\left( d+2-\bar{\eta}\right)
/2}\rho _{0}\right) .  \label{GEPORsc}
\end{equation}
The initial infection density $\rho _{0}$ plays here the role of an ordering
field. In the infinite time limit at criticality, when $\tau =0$, the order
parameter $\bar{\rho}_{\text{stat}}=\bar{\rho}\left( t\rightarrow \infty
\right) $ goes to zero with $\rho _{0}$ as
\begin{eqnarray}
\bar{\rho}_{\text{stat}}\left( \tau =0,\rho _{0}\right)  &\propto &\rho
_{0}^{\;1/\bar{\delta}},\qquad   \nonumber \\
\text{with}\qquad \bar{\delta} &=&\frac{d+2-\bar{\eta}}{d-2+\bar{\eta}}=
\frac{d+\sigma }{d-\sigma }.  \label{GEPkrit}
\end{eqnarray}
Below threshold, that is $\tau >0$, the order parameter is linear in $\rho
_{0}$ with a coefficient that diverges as $\tau \rightarrow 0^{+}$:
\begin{eqnarray}
\bar{\rho}_{\text{stat}}\left( \tau >0,\rho _{0}\right)  &\propto &\rho
_{0}\tau ^{-\bar{\gamma}},\qquad   \nonumber \\
\text{with}\qquad \bar{\gamma} &=&\left( 2-\bar{\eta}\right) \nu =\sigma \nu
\nonumber \\
&=&1+\frac{{\bar{\varepsilon}}}{4\sigma }{+O}\left(
\bar{\varepsilon}
^{2}\right) .
\end{eqnarray}
Lastly for $\tau <0$, the order parameter is independent of $\rho _{0}$ in
the limit $\rho _{0}\rightarrow 0$, and goes to zero with $\tau $ as
\begin{equation}
\bar{\rho}_{\text{stat}}\left( \tau <0\right) \propto \left| \tau \right| ^{
\bar{\beta}},  \label{GEPsup}
\end{equation}
where the order parameter exponent is given by
\begin{eqnarray}
\bar{\beta} &=&\nu \frac{\left( d-2+\bar{\eta}\right) }{2}=\nu \frac{\left(
d-\sigma \right) }{2}  \nonumber \\
&=&1-\frac{{\tilde{\varepsilon}}}{4\sigma }{+O}\left(
\bar{\varepsilon}
^{2}\right) .  \label{GEPbeta}
\end{eqnarray}
At the critical dimension $d=3\sigma $, the scaling behavior of
$\bar{\rho}_{
\text{stat}}$
\begin{eqnarray}
\bar{\rho}_{\text{stat}}(\tau  &=&0,\rho _{0})\propto \rho _{0}^{\;1/2}
\nonumber \\
\bar{\rho}_{\text{stat}}(\tau  &>&0,\rho _{0})\propto \rho _{0}\frac{\ln
^{1/4}\tau }{\tau }  \nonumber \\
\bar{\rho}_{\text{stat}}(\tau  &<&0)\propto \left| \tau \right| \ln
^{1/4}\left( \frac{1}{|\tau|} \right) .  \label{GEPlogK}
\end{eqnarray}
is mean-field like with logarithmic corrections. At criticality we find from 
(\ref{GEPORsc}) the scaling behavior
\begin{equation}
\bar{\rho}\left( t,\rho _{0}\right) =\rho _{0}^{\;\left( d-\sigma \right)
/\left( d+\sigma \right) }{\cal F}\left( t\rho _{0}^{\;2z/\left( d+\sigma
\right) }\right)   \label{GEPtime}
\end{equation}
with a universal scaling function
\begin{equation}
{\cal F}\left( x\right) \propto \left\{
\begin{array}{ll}
x^{\sigma /z} & \qquad \text{for\quad }x\ll 1 \\
1 & \qquad \text{(exponentially) for\quad }x\to \infty.
\end{array}
\right.   \label{GEPtsc}
\end{equation}
Note that the ``static'' exponents $\bar{\beta}$, $\bar{\gamma}$,
$\bar{
\delta}$, $\bar{\eta}$ (and $\nu $) correspond to well known undirected
percolation exponents but for long-range connectivity. They were already
given in \cite{PrLu76,Am76}.

To study the spread of the infection by computer simulations one may
investigate the response function $\bar{\chi}({\bf x},t)=\langle S\left(
{\bf x},t\right) \tilde{s}(0,0)\rangle $ which describes the density of the
immune percolating individuals at the time $t$ caused by an infection at $t=0
$, ${\bf x}=0$. Its scaling behavior is given by
\begin{equation}
\bar{\chi}(x,t,\tau )=b^{-\left( d-2+\tilde{\eta}\right) }\bar{\chi}\left(
b^{-1}{\bf x},b^{-z}t,b^{1/\nu }\tau \right) .  \label{GEPchi}
\end{equation}
We read off the correlation lengths for space and time and the corresponding
exponents as
\begin{eqnarray}
\xi  &\propto &\left| \tau \right| ^{-\nu }\qquad   \nonumber \\
\text{with}\qquad \nu  &=&\frac{1}{\sigma }+\frac{{\bar{\varepsilon}}}{
4\sigma ^{2}}{+O}\left( \bar{\varepsilon}^{2}\right) ,  \label{GEPxi} \\
\xi _{t} &\propto &\left| \tau \right| ^{-\nu _{t}}\qquad   \nonumber \\
\text{with}\qquad \nu _{t} &=&z\nu =1+\frac{{\bar{\varepsilon}}}{16\sigma }{
+O}\left( \bar{\varepsilon}^{2}\right) .  \label{GEPxit}
\end{eqnarray}
At criticality, $\tau =0$, we have
\begin{equation}
\bar{\chi}\left( {\bf x},t\right) =\left| {\bf x}\right| ^{-\left( d-2+\bar{
\eta}\right) }{\cal F}_{\chi }\left( \left| {\bf x}\right| {\bf /}
t^{1/z}\right)   \label{GEPchicr}
\end{equation}
with an universal scaling function ${\cal F}\left( x\right) .$ The Fourier
transformed susceptibility at the critical point scales as
\begin{equation}
\chi \left( {\bf q},\omega \right) =q^{-2+\bar{\eta}-z}\widetilde{{\cal F}}
\left( \omega {\bf /}q^{z}\right) \propto q^{-\sigma -z}.  \label{GEPchf}
\end{equation}
As the last scaling exponent that can be deduced in the usual way from the
given critical exponents we present the fractal dimension of the percolation
clusters of the removed individuals:
\begin{equation}
d_{f}=d-\frac{\bar{\beta}}{\nu }=\frac{d+\sigma }{2}.  \label{GEPfr}
\end{equation}
We note that the value of the exponent $\nu _{t}$ (\ref{GEPxit}) that we
have found is different from that given by Grassberger \cite{Grassberger}.
The values of all the exponents change {\em continuously }at the stability
boundary $\sigma =2\left( 1-\alpha \right) =2-\bar{\eta}_{GEP}=2+\varepsilon
/21+O\left( \varepsilon ^{2}\right) $ to their short-range undirected
percolation values.

\section{Conclusions}

\label{Concl}

Epidemic processes are growth models for phenomena arising abundantly in
nature. We have shown by imposing a L\'{e}vy flight type of infection
spreading that new long-range determined universality classes come into
play. We were able to characterize the universality classes by determining
the critical exponents to first order in an $\varepsilon $-expansion around
their upper critical dimension. There exist exact relationships between the
exponents, and some critical exponents are given exactly as functions of
spatial dimension and the exponent characterizing the long-range tail of the
L\'{e}vy flight infection. Besides, we have been able to build a
renormalization group flow that possess a fixed point structure that allows
to describe short-range and long-range infection in the same and to pass
{\em continuously} from one behavior to another by varying the L\'{e}vy
flight exponent. Because the anomalous susceptibility exponent, the analog
of the Fisher exponent in critical equilibrium phenomena, is negative here,
{\em the continuous} crossover between long- and short-range behavior arises
at a L\'{e}vy exponent {\em greater} than $2$. We hope that this work
triggers more simulational work on this subject.

\acknowledgments
This work has been supported in part by the SFB 237 (,,Unordnung und gro\ss
e Fluktuationen``) of the Deutsche Forschungsgemeinschaft.

\label{Refer}




\end{document}